\documentclass[final,5p,times,twocolumn]{elsarticle}

\usepackage{lineno,hyperref}
\usepackage{caption}
\usepackage{verbatim}

\journal{Journal of Non-Newtonian Fluid Mechanics}

\begin{document}

\begin{frontmatter}

\title{Steady streaming flows in viscoelastic liquids}

\author{Giridar Vishwanathan}
\author{Gabriel Juarez\corref{mycorrespondingauthor}}
\address{Department of Mechanical Science and Engineering, University of 
Illinois at Urbana-Champaign, Urbana, Illinois, 61801, USA}
\cortext[mycorrespondingauthor]{Corresponding author}
\ead{gjuarez@illinois.edu}

\begin{abstract}

We discuss experimental investigations on steady streaming flows of dilute and 
semi-dilute polymer solutions in microfluidic devices. The effect of non-Newtonian 
behavior on steady streaming for different model fluids is determined by characterizing 
the evolution of the inner streaming layer as a function of oscillation frequency using 
particle tracking velocimetry. We find that steady streaming velocity profiles in
constant-viscosity elastic liquids are qualitatively similar to those in Newtonian 
liquids. Steady streaming velocity profiles in elastic liquids with strong shear 
thinning, however, display two unique features: (i) a non-monotonic evolution of the
inner streaming layer with increasing frequency, first growing then decreasing in width, 
and (ii) a clear asymmetry in the flow profile at high frequencies.

\end{abstract}

\begin{keyword}
steady streaming, viscoelastic, microfluidics
\end{keyword}

\end{frontmatter}


\section*{Haiku}

Endless vorticies

A slippery solution

I see within them

\section{Introduction}

Steady streaming broadly refers to the steady, inertially rectified flow that occurs 
in the presence of a primary oscillatory flow \cite{Andrade,Riley97,Riley}. 
The phenomenon of steady streaming is a viable tool for microfluidic applications ranging 
from particle sorting \cite{Wang,Thameem}, mixing \cite{Sritharan,Lutz,Ahmed}, and 
trapping \cite{Marmottant,Lutz02,Lieu,Yazdi}. Therefore, there is considerable interest 
in applying these flows to biological systems as a purely hydrodynamic method of 
manipulation \cite{Wiklund, Friend}. While non-Newtonian liquids are frequently 
encountered in biological systems, steady streaming in non-Newtonian liquids remains 
poorly understood, particularly in the context of micro-scale flows where there is 
maximum potential for application. 

The classical problem of steady streaming around a cylinder of size $a$, oscillating 
with a frequency $f$, or angular frequency $\omega = 2 \pi f$, and a far-field amplitude $s$ 
in a Newtonian liquid is theoretically well-understood \cite{Riley,Wang68,Holtsmark,Bertelsen},
and thus provides a convenient baseline for studying steady streaming in non-Newtonian 
liquids. For a Newtonian liquid of kinematic viscosity $\nu$, the streaming flow around a 
cylinder is caused by the rectified advection of vorticity from the curved surface. The 
streaming flow field typically manifests as two distinctly separated layers having 
four-fold symmetry with four vortices in each. The streamlines in the inner, viscous 
driving layer depend only on the Stokes boundary layer width ($\sqrt{\nu/\omega}$) for 
a particular cylinder. The vortices in the outer, inertially driven layer rotate in the 
opposite sense and the flow profile is determined by the distance to the outer boundary. 
The inner layer grows with decreasing frequency and becomes very large when 
$\sqrt{\nu/\omega}\sim a$ \cite{raney}. The pathlines in both layers are independent of 
oscillation amplitude provided it is small, that is, $s/a\ll1$. The streaming velocity, 
however, scales uniformly in magnitude with the oscillation amplitude as $s^2$.

Dramatic changes in the steady streaming flow field of dilute polymer solutions around 
a cylinder oscillating at low frequencies ($10 \leq f \leq 100$ Hz) were experimentally 
observed \cite{Chang74}. In contrast to the Newtonian case, the inner streaming layer 
in non-Newtonian liquids was found to grow in size with increasing frequency, eventually 
displacing the outer driven vortices completely and giving the appearance of a reversed 
flow \cite{Chang79}. The growth of the inner streaming layer was theoretically described
using viscoelastic constitutive models \cite{frater,Chang77,James,Bohme}, however, few 
studies have investigated the link between the steady streaming flow and the measured 
rheological properties of non-Newtonian liquids. One such study used steady streaming as 
a rheological tool to characterize drag-reduction in dilute polymer solutions 
\cite{Vlassopoulos}. While a qualitative correlation between the drag reduction performance 
and the liquid relaxation time was observed, inconsistencies in the measured relaxation 
times from steady streaming differed by two orders of magnitude when compared to estimates 
from steady shear rheology. This key observation implied that the longest relaxation 
time may not be the important characteristic time when considering steady streaming in 
non-Newtonian liquids.

The effectiveness of steady streaming as a microrheological tool has been reconsidered 
recently and demonstrated for low-viscosity Newtonian liquids in microfluidic devices
\cite{Vishwanathan}. In this work, we experimentally investigate steady streaming of 
non-Newtonian liquids in microfluidic devices. A major advantage of the microscale 
approach is that higher frequencies can be accessed due to lower system inertia arising 
from small length scales $\mathcal{O}(100 \ \mu\textrm{m})$. Further, smaller cylinder 
radii enable larger strains and improved spatial-temporal resolution of the rotational 
inner layer where the liquid rheology is critical, even though comparatively higher 
frequencies are used. The motivation for our study is three-fold. First, 
to make quantitative observations on the inner streaming layer of non-Newtonian liquids 
in the context of microfluidics where they stand to be most likely encountered in 
contemporary applications. Second, to explore the relationship between the bulk rheology 
and molecular properties of model dilute and semi-dilute polymer solutions with the 
observed streaming flows, and third, to elucidate the possible mechanism by which the 
non-Newtonian behavior manifests in streaming.

\section{Experimental methods}

\subsection{Microfluidic steady streaming}

Experiments were performed in microfluidic devices molded in PDMS, consisting of a 
straight channel $20$ mm long, $5$ mm wide, and $200 \ \mu$m tall. A fixed cylindrical 
post with radius of $a = 100 \ \mu$m was manufactured at the center of the straight 
channel. An oscillatory flow field, $U(t)=s\omega \cos(\omega t)$, was setup in the 
channel through an external oscillating pressure signal generated using an 
electro-acoustic transducer, or loudspeaker, over a range of frequencies, 
$50 \leq f \leq 1000$ Hz. The oscillation amplitude was independently controlled over 
a range, $5 < s < 50 \ \mu$m, such that the non-dimensional amplitude 
($\epsilon = s/a \ll 1$) is small. For a given data set, the value of $\epsilon^2 \omega$ 
was maintained constant within $\pm \ 20 \%$ of a mean value which lies in a range of 
$60-180$ rad s$^{-1}$, depending on the viscosity of the liquid used. Note that 
$\epsilon^2 \omega$ must be large enough for the streaming velocity to be unaffected 
by Brownian diffusion, and small enough that $\epsilon<0.5$, even at the lowest 
frequency. The corresponding maximum streaming velocities measured for this range of 
$\epsilon^2 \omega$ is between $50-500$ $\mu$m/s, for a particular data set. The
Reynolds number, $\textrm{Re} = \omega a^2/\nu$, and the streaming Reynolds number,
$\textrm{Re}_\textrm{s} = \omega s^2/\nu$, correspond to a range of 
$2 \leq \textrm{Re} \leq 50$ and $0.1 \leq \textrm{Re}_\textrm{s} \leq 1$, 
respectively, for deionized water ($\nu = 0.949\times 10^{-6}$ m$^2$/s) over the 
entire frequency range investigated here. This range of $\textrm{Re}$ results in an 
inner streaming layer that is large compared to the cylinder radius and enables high 
resolution velocimetry of the inner layer.

Polystyrene tracer particles, $0.93 \ \mu$m in diameter, were observed at the mid-height 
of the straight channel using bright field microscopy at $20\times$ magnification (depth 
of focus $\approx 3\ \mu$m). Images were acquired using a scientific CMOS camera where the
sampling frequencies are much greater than (high-speed), or perfect divisors (stroboscopic) 
of the oscillatory flow frequency. High-speed imaging provided high fidelity observation 
of the oscillatory flow component and measurement of the oscillation amplitude $s$, 
while stroboscopic imaging was used to measure the streaming velocity fields. 
Experiments were performed at room temperature, maintained at $20 \ ^{\circ}$C.

\begin{table}[b]
\begin{center}
\caption{Properties of the various polymer molecules and aqueous polymer solutions 
used in this study.}
\label{table:solutions}
\begin{tabular}{lcccc}
\hline
\begin{tabular}[c]{@{}c@{}} Polymer \\ solution\end{tabular} & 
\begin{tabular}[c]{@{}c@{}} $M_w$   \\ {[}MDa{]}\end{tabular} & 
\begin{tabular}[c]{@{}c@{}} $c^*$   \\ {[}ppm{]}\end{tabular} & 
\begin{tabular}[c]{@{}c@{}} $c$     \\ {[}ppm{]}\end{tabular} & 
\begin{tabular}[c]{@{}c@{}} $c/c^*$ \\ {}\end{tabular}  
 \\ \hline
XG    & 2.7 & 1600 & 400  & 0.25 \\ \hline
XG    &     &      & 1000 & 0.63 \\ \hline
PAA   & 6   & 550  & 4000 & 7.3  \\ \hline
hPAA  & 18  & 200  & 20   & 0.1  \\ \hline
hPAA  &     &      & 50   & 0.25 \\ \hline
hPAAs &     &      & 500  & 2.5  \\ \hline
\end{tabular}
\bigskip
\caption*{Here, $M_w$ is polymer molecular weight, $c^*$ and $c$ are overlap and solute concentrations, respectively.}
\end{center}
\end{table}

\subsection{Polymer solutions}

Solutions were prepared by step-wise dissolution of different polymers into deionized 
(DI) water with gentle agitation at $60$ rpm for $4-6$ hours. After initial dispersion, 
the required final concentration was achieved by successive dilution with DI water. The 
polymers used in this study were xanthan gum (XG, $2.7 \times 10^6$ MW, Sigma Aldrich G1253),
non-ionic polyacrylamide (PAA, $6 \times 10^6$ MW, Polysciences 02806), and 
polyacrylamide-acrylate co-polymer with a $30\%$ degree of hydrolysis (hPAA, $18 \times 10^6$ 
MW, Polysciences 18522). In addition to the above aqueous solutions, salinated 
polyacrylamide-acrylate (hPAAs) was prepared by dissolving hPAA in $0.5$ M NaCl solution. 
The properties of the polymer molecules and the polymer solutions are listed in Table 
\ref{table:solutions}. Estimates of the overlap concentration $c^*$ for these model 
solutions have been reported in literature \cite{Kulicke,Turkoz,Francois79}. Here, all 
solution concentrations are reported in parts-per-million (ppm) by weight, that is, 1 
ppm = $10^{-6}$ gram of solute per gram of solvent.

\subsection{Bulk rheology}

\begin{figure}
\centering
	\includegraphics[width=\linewidth]{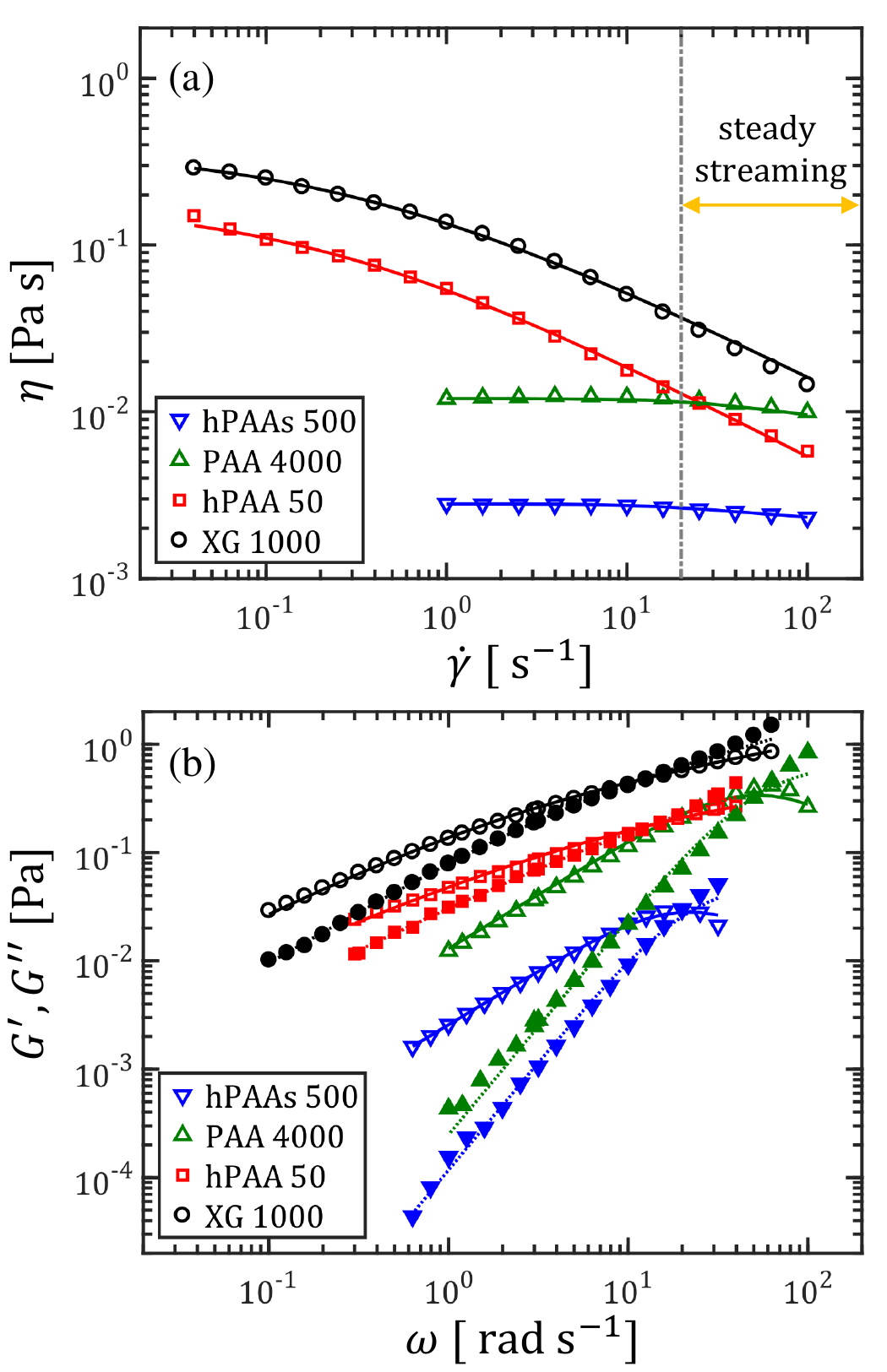}
\caption{
(a) Steady state shear viscosity measurements of dilute and semi-dilute polymer solutions
used in this study and respective fits (solid line). The yellow arrow shows the mean range
of shear rates encountered in experiments. (b) Storage (closed symbols) and loss (open
symbols) moduli from oscillatory shear rheology of polymer solutions and model fits shown 
as solid and dashed lines, respectively. The model parameters used are listed in Table
\ref{table:fits}.
}
	\label{fig:figone}
\end{figure}

The steady shear viscosity ($\eta$) versus shear rate ($\dot{\gamma}$), and the storage
($G^{\prime}$) and loss ($G^{\prime\prime}$) moduli versus angular frequency ($\omega$) 
for some polymer solutions are shown in Figure \ref{fig:figone}. We characterize all 
liquids using a temperature-controlled cone-and-plate geometry rheometer (strain-controlled
Ares G2, TA Instruments) at 25 $^\circ$C. We find strong shear thinning in $1000$ ppm 
xanthan gum (XG $1000$) and $50$ ppm polyacrylamide-acrylate (hPAA $50$) solutions, 
shown in Figure \ref{fig:figone}(a). At lower concentration solutions of XG $400$ and 
hPAA $20$ (not shown), shear thinning behavior is decreased. Shear thinning is minimal  
for the $4000$ ppm non-ionic polyacrylamide (PAA $4000$) and $500$ ppm salinated 
polyacrylamide-acrylate (hPAAs $500$) solutions, which can be considered Boger fluids, 
or constant-viscosity elastic liquids \cite{Boger,DFJames}. 

The steady shear viscosity for the solutions shown in Figure \ref{fig:figone}(a) are 
modelled using a Carreau model \cite{bird}, determined by a zero-shear viscosity $\eta_0$, 
infinite-shear viscosity $\eta_{\infty}$ (here, assumed to be solvent viscosity), power-law
thinning index $n$ and a relaxation time $\lambda_{Cr}$. The dynamic moduli of PAA 
$4000$ and hPAAs $500$ shown in Figure \ref{fig:figone}(b) are described well by the 
Maxwell model \cite{bird} and characterized by a viscosity $\eta_M$, and relaxation 
time $\lambda_M$. The dynamic moduli for XG $1000$ and hPAA $50$ on the other hand, 
are modelled well using a fractional Maxwell model (FMM) \cite{jaishankar} described 
by two dimensionless exponents: $0\leq\beta\leq0.5$ and $\beta\leq\alpha\leq1$; along 
with two stiffness quasi-properties $\mathbb{G}$ and $\mathbb{V}$. The model based fits 
are shown in Figure \ref{fig:figone} as solid or dashed lines, while the model parameters 
are listed in Table \ref{table:fits}. 

\begin{table}
\begin{center}
\caption{ Rheological fit parameters 
used in this study}
\label{table:fits}
\begin{tabular}{lccccc}
\hline
\begin{tabular}[c]{@{}c@{}} \\Carreau \end{tabular} & 
\begin{tabular}[c]{@{}c@{}}$\eta_0$\\ {[}Pa s{]}\end{tabular} & 
\begin{tabular}[c]{@{}c@{}}$\eta_\infty$\\ {[}Pa s{]}\end{tabular} & 
\begin{tabular}[c]{@{}c@{}}$\lambda_{Cr}$\\{[}s{]}\end{tabular} & 
\begin{tabular}[c]{@{}c@{}}\\ $n$\end{tabular} & 
\begin{tabular}[c]{@{}c@{}}\\ $R^2$\end{tabular}  
\\ \hline
hPAA 50  &  0.119  & 0.001  & 4.96 & 1.51 & 0.993      \\ \hline
XG 1000 & 0.26  & 0.001 & 3.37 & 1.48 & 0.988       \\ \hline
hPAAs 500 & 0.0028  & 0.001 & 0.074 & 1.15 & 0.996       \\ \hline
PAA 4000 & 0.012  & 0.001 & 0.046 & 1.16 & 0.999       \\ \hline
\begin{tabular}[c]{@{}c@{}}\\ Maxwell\end{tabular} & 
\begin{tabular}[c]{@{}c@{}}$\eta_M$\\ {[}Pa s{]}\end{tabular} & 
\begin{tabular}[c]{@{}c@{}}$\lambda_M$\\ {[}s{]}\end{tabular} & 
\begin{tabular}[c]{@{}c@{}}{    }\\ \end{tabular} & 
\begin{tabular}[c]{@{}c@{}}{    }\\ \end{tabular} & 
\begin{tabular}[c]{@{}c@{}}\\$R^2$\end{tabular}  
\\ \hline
PAA 4000  & 0.013   & 0.019  & & & 0.987      \\ \hline
hPAAs 500 & 0.0026  & 0.046 & & & 0.973       \\ \hline
\begin{tabular}[c]{@{}c@{}} \\FMM \end{tabular} & 
\begin{tabular}[c]{@{}c@{}}$\mathbb{V}$\\ {[}Pa s$^\alpha${]}\end{tabular} & 
\begin{tabular}[c]{@{}c@{}}$\mathbb{G}$\\ {[}Pa s$^{ \,\beta}${]}\end{tabular} & 
\begin{tabular}[c]{@{}c@{}}\\$\alpha$\end{tabular} & 
\begin{tabular}[c]{@{}c@{}}\\ $\beta$\end{tabular} & 
\begin{tabular}[c]{@{}c@{}}\\ $R^2$\end{tabular} 
\\ \hline
hPAA 50  &  0.071  & 0.22  & 0.75 & 0.27 & 0.991      \\ \hline
XG 1000 & 0.23  & 0.39 & 0.86 & 0.34 & 0.996       \\ \hline

\end{tabular}

\bigskip
\caption*{The table shows the model parameters used to obtain the fits shown in Figure 
\ref{fig:figone}. $R^2$ is a residual quantifying the agreement between the fit 
and the experimental data.}
\end{center}
\end{table}

\begin{table} [b]
\begin{center}
\caption{Relaxation time estimates from different models}
\label{table:relaxationtimes}
\begin{tabular}{lccccc}
\hline
\begin{tabular}[c]{@{}c@{}}Polymer\\ solution\end{tabular} & 
\begin{tabular}[c]{@{}c@{}}$\lambda_Z$\\ {[}ms{]}\end{tabular} & 
\begin{tabular}[c]{@{}c@{}}$\lambda_{co}$\\ {[}ms{]}\end{tabular} & 
\begin{tabular}[c]{@{}c@{}}$\lambda_{Cr}$\\ {[}ms{]}\end{tabular} & 
\begin{tabular}[c]{@{}c@{}}$\lambda_{FMM}$\\ {[}ms{]}\end{tabular} 
 \\ \hline
XG 1000   & - & 79 & 3370 & 362 \\ \hline
PAA 4000  & 2.7 & 17 & 46 & - \\ \hline
hPAA 50 & -  & 83  & 4960 & 94.7 \\ \hline
hPAAs 500 & 11.2 & 50 & 74  & - \\ \hline
\end{tabular}
\bigskip
\caption*{Above, $\lambda_Z$ is from Zimm theory, $\lambda_{co}$ is from the crossover 
of dynamic moduli, $\lambda_{Cr}$ is from the Carreau model, and $\lambda_{FMM}$ is 
from the fractional Maxwell model. The relaxation times may be compared to 
$\lambda_{ref}=0.4$ ms, for which $\lambda_{ref} \omega = 1 $ at $400$ Hz.}
\end{center}
\end{table}

\begin{figure*}
\centering
	\includegraphics[width=\linewidth]{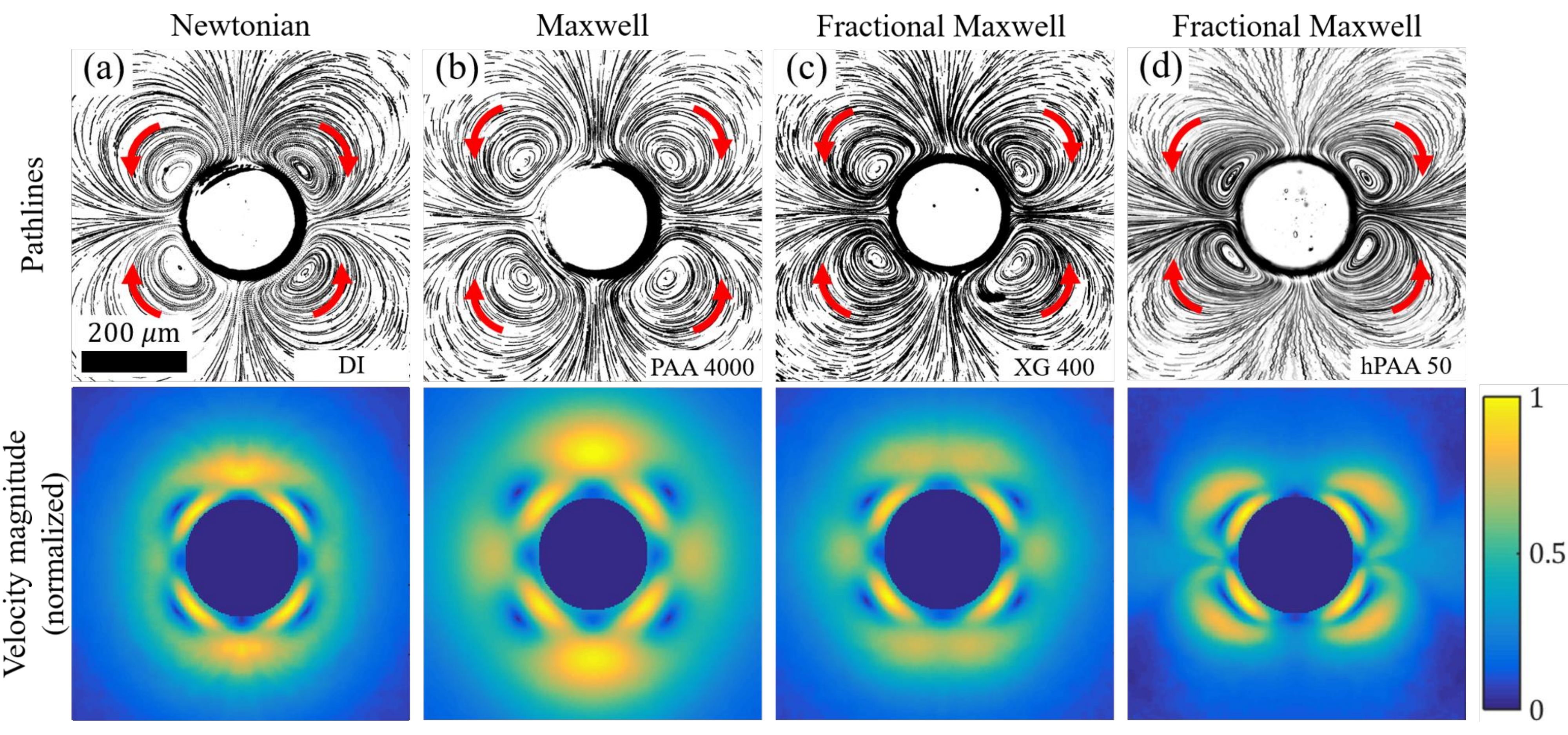}
\caption{
Steady streaming profiles in Newtonian and non-Newtonian liquids at an 
oscillation frequency of $600$ Hz around a cylinder with radius of $100 \ \mu$m. 
(Top row) Pathlines of tracer particles for (a) DI water and $\epsilon=0.07$, (b) 
$4000$ ppm non-ionic polyacrylamide and $\epsilon=0.08$, (c) $400$ ppm xanthan gum 
and $\epsilon=0.15$, and (d) $50$ ppm hydrolyzed polyacrylamide and $\epsilon=0.19$.
(Bottom row) Corresponding steady streaming velocity magnitude field, normalized 
by the maximum streaming velocity, obtained from particle tracking velocimetry.
High-velocity regions, light (or yellow) regions of the colormap, are located near 
the cylinder boundary.
}
	\label{fig:figtwo}
\end{figure*}

A relaxation time can also be inferred from the crossover angular frequency 
($\omega_{co}$) at which the storage and loss moduli are equal and determined by
$\lambda_{co}=1/\omega_{co}$. A more rigorous estimate of the relaxation time from the 
model parameters are $\lambda_M\approx\lambda_{co}$ for a Maxwell liquid and
$\lambda_{FMM}=(\mathbb{V}/\mathbb{G})^{1/(\alpha-\beta)}$ for a fractional Maxwell 
liquid. Yet another estimate of the relaxation time for the constant-viscosity polymer 
solutions (dilute and semi-dilute unentangled) can be obtained from molecular 
considerations using the Zimm formula \cite{zimm}:
\begin{equation}
    \lambda_Z = \frac{\eta_s [\eta]_0 M_w}{2.37RT} \ ,
\end{equation}
where $\eta_s$ is the solvent viscosity, $[\eta]_0$ is the intrinsic shear viscosity at 
zero shear rate, $R$ is the universal gas constant, and $T$ is the temperature. Here, 
we take $\eta_s=\eta_{\infty}=1\times 10^{-3}$ Pa s (DI water) and $[\eta]_0$ was 
approximated for the solutions used based on the zero-shear viscosity and solvent 
viscosity. The various estimates of relaxation times from the different methods are 
listed in Table \ref{table:relaxationtimes}. Note that the Zimm theory cannot be 
applied to polyelectrolyte solutions such as XG and hPAA in the absence of salts 
\cite{Turkoz}.

\section{Results}

The comparison between steady streaming in Newtonian and non-Newtonian liquids at an 
oscillation frequency of $600$ Hz is shown in Figure \ref{fig:figtwo}. The particle 
pathlines (top row) are generated from minimum intensity projections of a sequence of 
stroboscopic images and the velocity magnitude fields (bottom row) are obtained from 
particle tracking velocimetry. For all cases, the quadrupolar topology of the streaming 
flow field is preserved, consisting of four identical vortices with distinct centers. 
Similarly, the rectified flow moves toward the cylinder along the axis of oscillation 
and away from the cylinder normal to the axis of oscillation.

The pathlines and velocity magnitude field for a Newtonian liquid, deionized water, 
are shown in Figure \ref{fig:figtwo}(a, top and bottom). High flow velocities are 
located in the regions between the cylinder surface and the eddy center, and near the
cylinder surface, perpendicular to the axis of oscillation. This latter difference, 
compared to the velocity along the axis, is because of increased interaction with the 
channel wall in the breadthwise direction as compared to the lengthwise direction.

The steady streaming pathlines and velocity magnitude field of a Maxwell liquid, 
PAA $4000$, are qualitatively similar to that of a Newtonian liquid, shown in Figure 
\ref{fig:figtwo}(b, top and bottom). The main difference, compared with DI water, is 
the eddy center distance from the cylinder surface, which is larger for PAA $4000$.
This larger eddy center distance indicates a higher viscosity for PAA $4000$, as was 
recently demonstrated \cite{Vishwanathan}, and is in agreement with steady state 
shear viscosity measurements shown in Figure \ref{fig:figone}(a).

\begin{figure}
\centering
	\includegraphics[width=\linewidth]{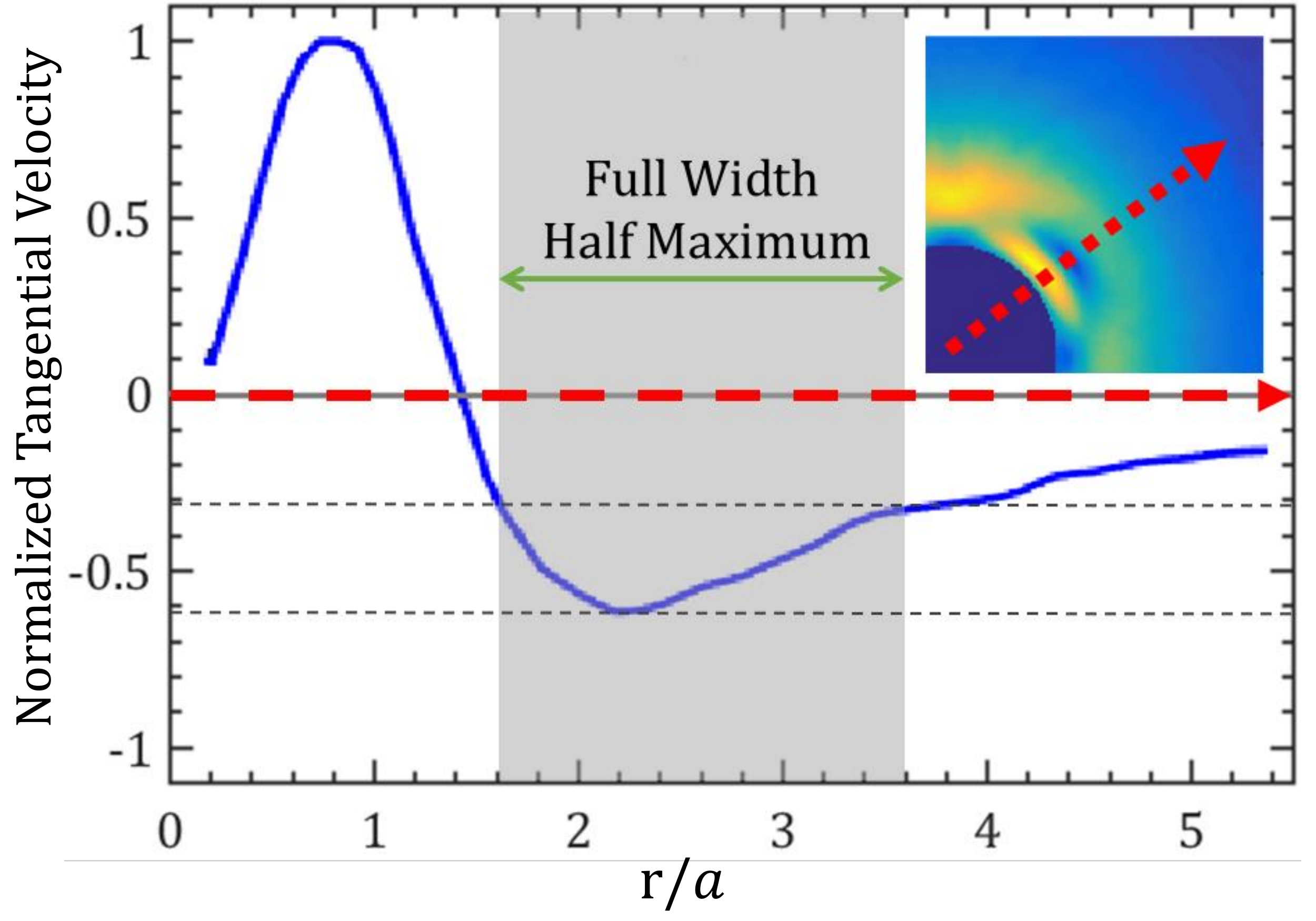}
\caption{
Representative steady streaming tangential velocity profile, normalized by 
the maximum streaming velocity, as a function of dimensionless radial position. This 
profile is taken along the transect starting from the cylinder surface and radially 
outward through the eddy center, as shown in the inset. The velocity is highest in 
magnitude near the cylinder surface and zero at the eddy center location, as it goes 
from positive to negative values. The full width at half maximum (FWHM) of the second 
(negative) peak is used to characterize the width of the inner layer and is annotated 
for DI water at an oscillation frequency of $100$ Hz.
}
	\label{fig:figthree}
\end{figure}

In the fractional Maxwell liquids, a qualitative difference is observed in particle 
pathlines and velocity magnitude fields. For XG $400$ solutions, the angular position 
of the eddy centers move closer to the axis of oscillation, shown in Figure 
\ref{fig:figtwo}(c, top). The velocity field has also observably changed with the 
velocity maximum, normal to the axis of oscillation, separated into two regions by a 
local minimum, shown in Figure \ref{fig:figtwo}(c, bottom). For hPAA $50$ solutions, 
a more prominent change is seen where the angular position of the eddy center as well 
as the velocity maximum are moved considerably towards the axis of oscillation, Figure
\ref{fig:figtwo}(d, top and bottom).

To quantify differences in the steady streaming profile for various liquids over
a range of frequencies, the normalized tangential velocity profile as a function 
of the dimensionless radial position ($\textrm{r}/a$) along a transect from the 
cylinder surface through the eddy center was examined. The transect is depicted 
by the red dashed arrow passing through the eddy center in Figure 
\ref{fig:figthree}(inset). A representative experimentally measured velocity profile 
for a Newtonian liquid is shown in Figure \ref{fig:figthree}. The particular parameter 
of interest is the full width at half maximum (FWHM) of the second peak, beyond the 
eddy center. The FWHM provides a measure of the size of the inner layer in this streaming 
regime where a distinctly visible separation between the inner and outer layers is absent. 
The FWHM is also preferred because of the insensitivity to inaccuracies in particle 
tracking close to the cylinder surface as well as inaccuracies in the measurement of 
oscillation amplitude ($s$).

\begin{figure}
\centering
	\includegraphics[width=\linewidth]{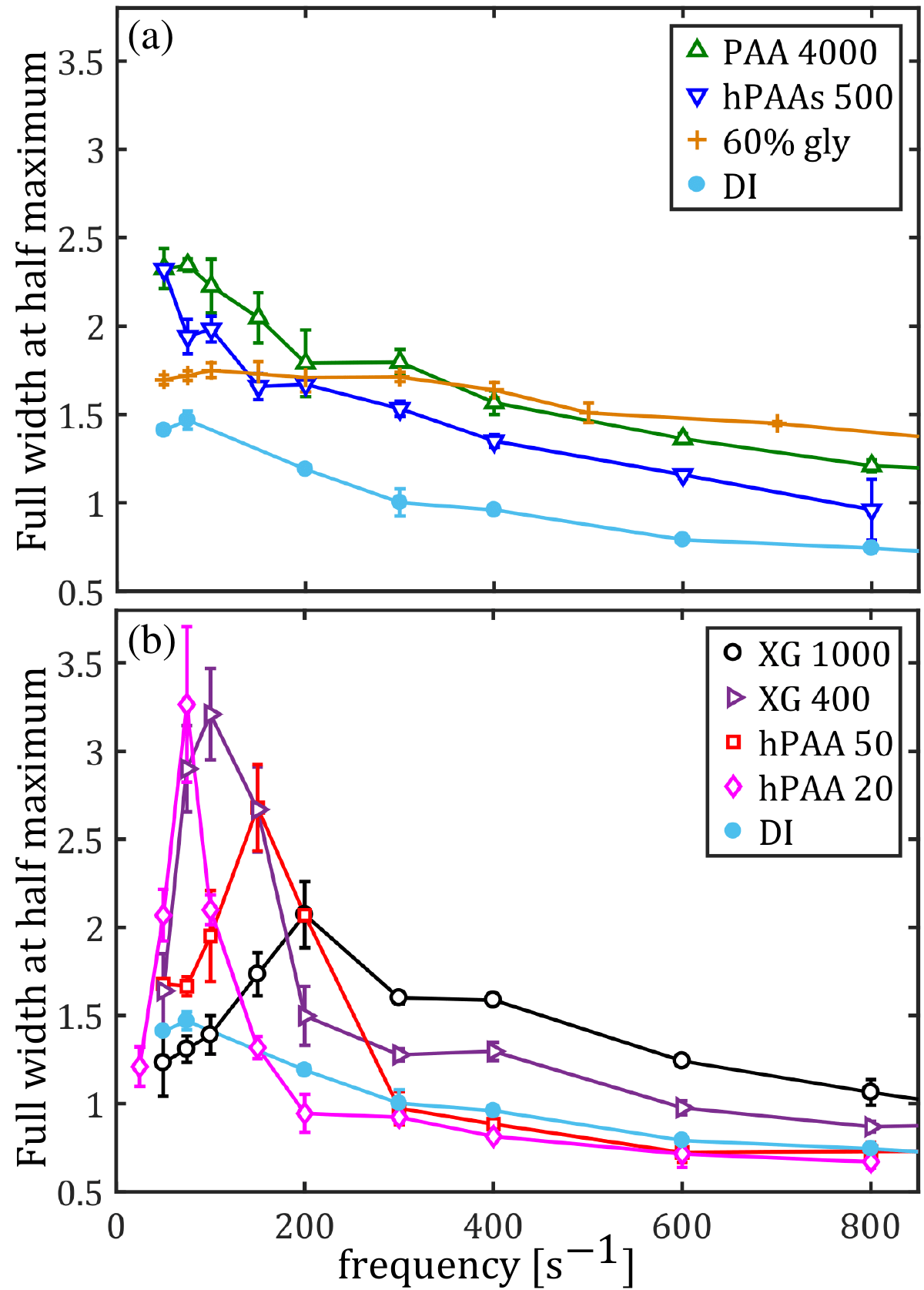}
\caption{
Evolution of inner streaming layer characterized by the full width at half 
maximum for Newtonian and non-Newtonian liquids as a function of oscillation frequency 
in microfluidic devices. (a) The FWHM of Newtonian liquids and Maxwell liquids exhibit
monotonic decrease with increasing oscillation frequency. (b) The FWHM of fractional 
Maxwell liquids, however, increases to a maximum value and then decreases with 
increasing oscillation frequency.
}
	\label{fig:figfour}
\end{figure}

The variation of the FWHM with frequency for Newtonian liquids, DI water and aqueous 
$60\%$ glycerol (w/w) with a dynamic viscosity of $5.9$ mPa s, and Maxwell liquids, 
PAA $4000$ and hPAAs $500$, is shown in Figure \ref{fig:figfour}(a). The Newtonian 
liquids exhibit slow monotonic decrease in the FWHM with increasing oscillation 
frequency. The Maxwell liquids show a nominally faster rate of decrease in the FWHM 
with increasing frequency, without any discernible features. 

This is in sharp contrast with the observations for fractional Maxwell liquids. All 
four polymer solutions exhibit distinct peaks in their FWHM with increasing oscillation 
frequency, shown in Figure \ref{fig:figfour}(b). For hPAA $50$ ppm solutions, 
the FWHM grows rapidly for oscillation frequencies of $75$ Hz to $150$ Hz, which is 
in agreement with previous observations \cite{Vlassopoulos}. However, the FWHM begins 
to decrease with further increasing frequency until its variation is similar to that 
of the solvent. Similar trends are observed for hPAA $20$ solutions, with the maximum 
FWHM observed at $75$ Hz and a subsequent rapid decrease to Newtonian-like behavior. 

The XG $1000$ and XG $400$ solutions also behave in a manner consistent with 
our observations for the hPAA solutions, exhibiting a distinct peak in the FWHM
followed by a decrease to Newtonian-like behavior with increasing frequency. The 
key difference between the two polymers is the rate at which they decrease at 
large frequencies, which is slower for XG solutions. This is in agreement with the 
qualitative observations where the pathlines and the corresponding velocity magnitude 
fields are not as distorted in XG as they are in hPAA, even at $600$ Hz, shown 
in Figure \ref{fig:figtwo}(c) and (d). Further increase in frequency is not found 
to significantly alter the velocity profile of hPAA $20$ and hPAA $50$ solutions, 
while XG $400$ shows a gradual approach towards velocity fields similar to those of 
hPAA $50$. This approach is slower still for XG $1000$ and therefore reflected 
completely by the evolution of the FWHM shown in Figure \ref{fig:figfour}.

\section{Discussion}

The significant departure from four-fold symmetry in the streaming flow profile 
observed for Fractional Maxwell liquids, as compared to the Newtonian case in 
Figure \ref{fig:figtwo}(top), is attributed to the nature of the oscillatory 
strain field and the shear thinning rheology of these liquids (see Figure 
\ref{fig:figone}(a)). At leading order, the rate of strain 
($\mathcal{O}(\epsilon\omega)$) in the vicinity of the cylinder is completely 
extensional along the axis of oscillation, and purely shearing perpendicular to it. 
The spatial-temporal mean rates of extension and shear obtained theoretically 
\cite{Holtsmark} for DI water with $\nu=10^{-6}$ m${^2}$/s and a cylinder radius 
of $100\ \mu$m are $0.15 \epsilon \omega$ and $0.45 \epsilon \omega$, respectively.
For the frequencies used in this study, the mean strain rates are in the range 
$50-200$ s$^{-1}$ for shear (shown in Figure \ref{fig:figone}(a)) and $15-60$ 
s$^{-1}$ in the case of extension. The corresponding maximum, instantaneous rates of 
extension and shear are $0.3 \epsilon \omega$ and $3.73 \epsilon \omega$, respectively. 
The relatively large prefactor for shear is due to its localization near the cylinder 
surface. As a result, the effective viscosity is considerably lower in the high shear 
region for liquids exhibiting shear thinning, and consequently asymmetry is observed 
in the streaming flow profile. Maxwell liquids, on the other hand, exhibit minimal 
shear thinning behavior in this range of shear rates. Therefore, the location of the 
eddy center and overall streaming flow profile does not differ significantly in 
comparison to Newtonian liquids.

The growth of the inner streaming layer for a cylinder oscillating at frequencies 
$f < 100$ Hz in dilute hydrolyzed polyacrlyamide solutions has been experimentally 
observed \cite{Vlassopoulos}. Although the growth of the inner layer at unity Deborah 
numbers ($De=\lambda\omega\approx1$) was analytically understood to be an effect of 
elasticity using viscoelastic models \cite{frater,James,Chang77}, the most relevant 
estimate of the relaxation time remains ambiguous. Different estimates of relaxation 
time for liquids such as hPAA $50$ and XG $1000$ that show inner layer growth, 
are listed in Table \ref{table:relaxationtimes}. The estimates vary considerably 
depending on the rheology data used, all of which yield $De>10$, even at $20$ Hz. 

In contrast, the experimental results presented here for Maxwell liquids indicate 
a decrease of the inner layer width, without the characteristic growth. While this 
observation is consistent with previous experiments, which report almost Newtonian 
behavior for similar liquids \cite{Vlassopoulos}, we note that $De>10$ even at $100$ 
Hz, regardless of the relaxation time estimate used from Table \ref{fig:figthree}. 
This lack of inner layer growth is unsurprising when we consider that coil-stretch 
transitions are theorized to require $\dot{\varepsilon} \lambda_Z >1$ in purely 
extensional flow \cite{DeGennes,Larson}. Estimating the maximum instantaneous rate 
of extension in our experiments ($\dot{\varepsilon}\leq0.3\epsilon\omega$) to be 
$30-100$ s$^{-1}$, the values of $\lambda_Z$ shown in Table 
\ref{table:relaxationtimes} do not satisfy the criterion. Hence, polymer stretching 
in the primary flow is likely insufficient for the growth of the inner streaming 
layer in Maxwell liquids. 

The decrease of the inner layer width following the initial growth in the fractional 
Maxwell liquids has not been reported before and is best explained by a competition 
between the effects of inertia and elasticity. This competition was suggested in the 
theoretical analysis for an oscillating sphere (see Fig. 4 and Fig. 6 in Ref. \cite{Bohme}) 
and quantified through the material functions $S=\rho a^2 \omega^2/G^{\prime\prime}$ 
and $V=G^{\prime}/G^{\prime\prime}$, which are analogs of the Reynolds number and 
Deborah number, respectively. Similar to theories for a cylinder, increasing $V$ while 
keeping $S$ fixed was found to increase the width of the inner streaming layer. 
Increasing $S$ for a fixed $V$, however, results in a set of cascading same sense 
vortices, of which, the innermost vortex decreases in width. Owing to the dependence 
of both $S$ and $V$ on $\omega$, it is difficult to experimentally achieve independent 
control. 

\begin{figure}
\centering
	\includegraphics[width=\linewidth]{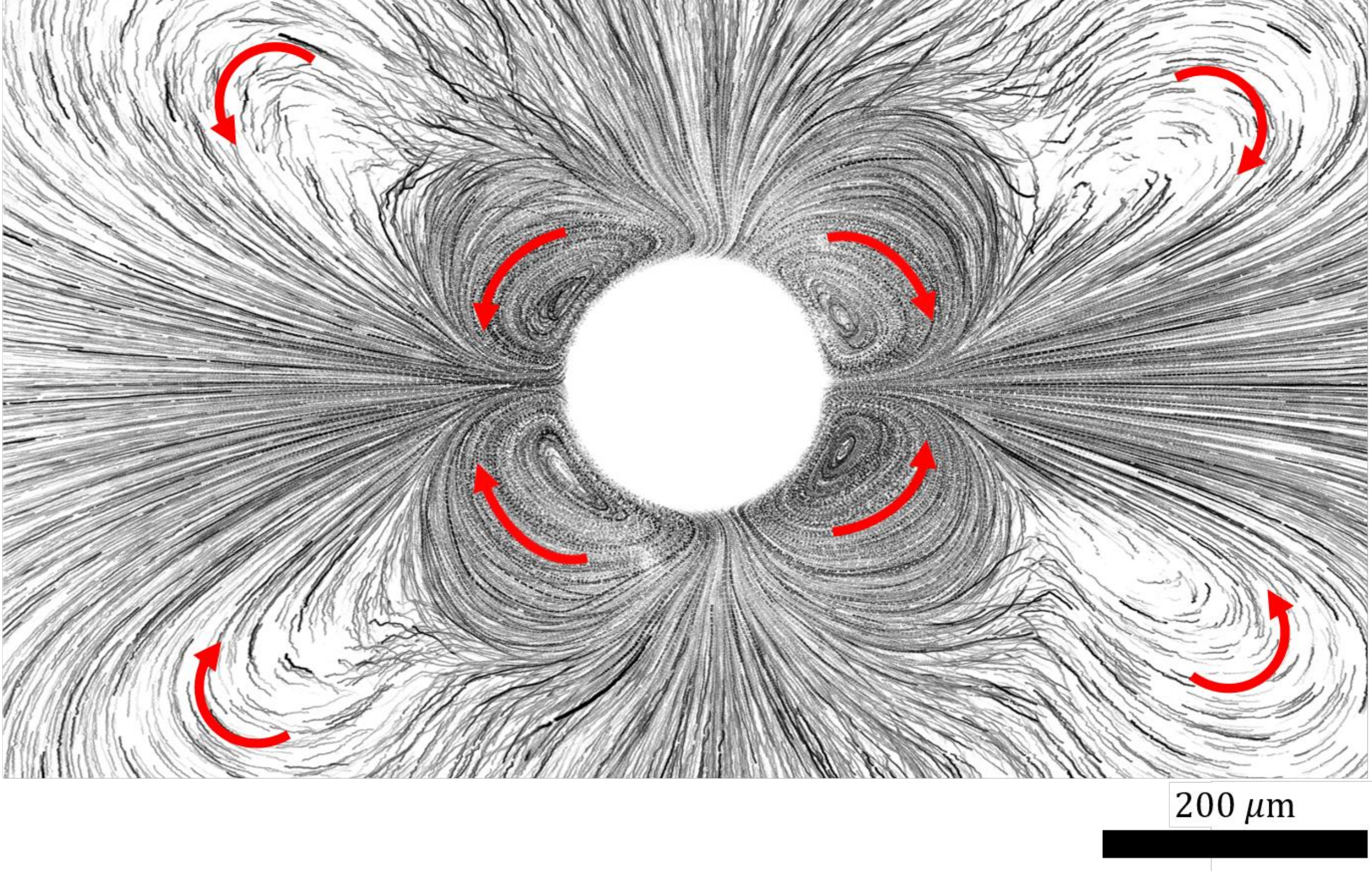}
\caption{
Pathlines of tracer particles illustrating experimental evidence of the predicted 
cascading inertio-elastic vortices. A set of secondary, same-sense vortices is shown 
for an hPAA $50$ polymer solution at an oscillation frequency of $800$ Hz and 
non-dimensional amplitude of $\epsilon=0.2$.
}
	\label{fig:figfive}
\end{figure}

\begin{figure}
\centering
	\includegraphics[width=\linewidth]{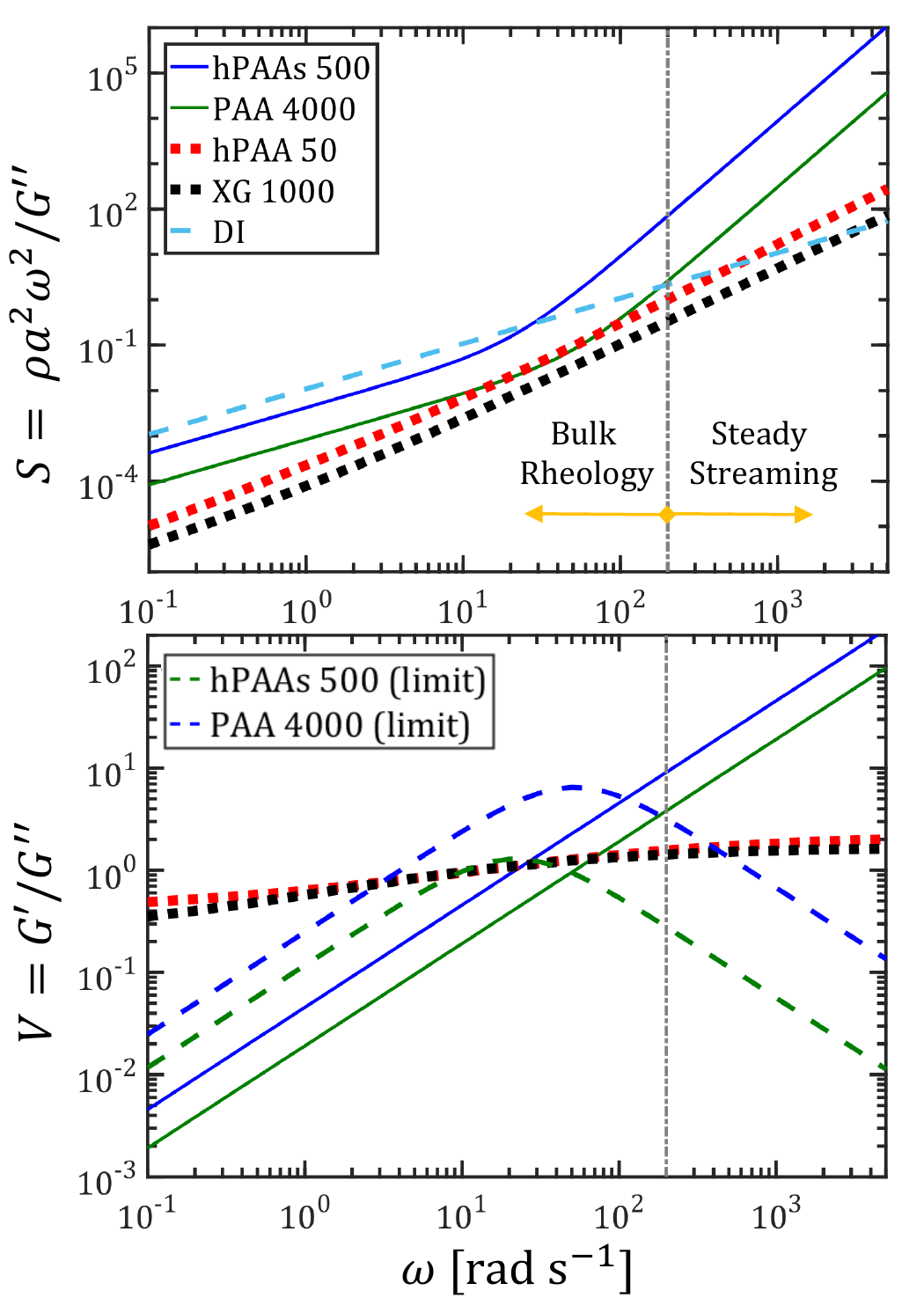}
\caption{
Variation of $S$ (top) and $V$ (bottom) with angular frequency based on the 
rheology fit models. Oscillatory rheology was performed in the frequency range to the 
left of the vertical line while streaming experiments were performed in the range of 
frequencies to the right.
}
	\label{fig:figsix}
\end{figure}

Regardless, the first experimental evidence of cascading same sense vortices for a 
fractional Maxwell liquid (hPAA $50$) at $800$ Hz is shown in Figure \ref{fig:figfive}.
An important implication of this observation is that, eventually, the effect of a 
growing $S$ dominates over that of a growing $V$. Indeed, we find that this is the case 
from Figure \ref{fig:figsix} which shows the variation of the material functions $S$ and 
$V$ with frequency for the models and fit parameters given in Table \ref{table:fits}. The 
values to the left of the vertical dotted line lie in the range of frequencies for which 
the oscillatory rheology was measured while those to the right are an extrapolation and 
lie in the frequency range for which streaming is studied. 

For fractional Maxwell liquids (dotted lines) and frequencies $f>100$ Hz, $V$ saturates 
in the range of $1.5<V<2$ while $S$ grows much faster to $10<S<100$. Therefore, there is 
quantitative concord with our experimental observations of growth and subsequent shrinking 
of the inner layer beyond $(V=1.5,S=50)$, and the corresponding values of $(V=2,S=50)$ shown
elsewhere \cite{Bohme} at which the cascading vortices manifest, albeit for a sphere. 
This suggests that a description of viscoelasticity in terms of $S$ and $V$ is more robust 
than one based on relaxation times.

For the Maxwell liquids, however, the material function $V$ increases rapidly when 
extrapolated to frequencies $f>20$ Hz (solid lines in Figure \ref{fig:figsix}(bottom).
Notwithstanding that the polymer chains are likely insufficiently stretched in the flow, 
it is worth noting that rapid growth of $V$ for a Maxwell liquid is unphysical because it 
requires an indefinitely decreasing $G^{\prime\prime}$ for an approximately constant 
$G^{\prime}$. The solvent viscosity (here $\eta_{\infty}=1$ mPa.s) provides a definite 
lower bound on the loss modulus, and hence, an upper bound for $V$ given by
$G^{\prime}/(\eta_{\infty} \omega)$ and shown in Figure \ref{fig:figsix}(bottom, dashed 
lines). Thus, for the Maxwell liquids, an increasing $\omega$ will eventually lead to a 
decrease in $V$ or, at least, a significant decrease in its growth with frequency. This 
is also true of $S$ for which is limited by the corresponding curve for DI water (top, 
dashed lines).

\section{Conclusions}

In this experimental work, we have studied the inner steady streaming layer of dilute and 
semi-dilute polymer solutions, having both Maxwell and fractional Maxwell oscillatory 
shear rheology. Qualitative differences in the velocity fields are observed for fractional 
Maxwell liquids, most prominently, an angular displacement of the eddy centers towards the
direction of oscillation and consequent asymmetry. The width of the inner streaming layer 
is characterized by the full width at half maximum of the velocity profile along the eddy 
center from the cylinder boundary. We find that the inner streaming layer width of fractional
Maxwell liquids exhibits a non-monotonic relationship with oscillation frequency; first 
increasing in width followed by a decrease in width with increasing frequency. This behavior 
is attributed to a competition between elasticity and inertia, with inertia dominating at 
high frequencies. Maxwell liquids, however, exhibit a monotonic decrease of the inner
streaming layer width with increasing frequency at a rate nominally faster than that 
observed for a Newtonian liquid. Even for oscillation periods considerably smaller than 
the molecular relaxation time, we do not observe any qualitative change in the flow 
profile as compared to a Newtonian liquid. This behavior is attributed to the insufficient
molecular stretching by the underlying oscillatory flow.

\section{Acknowledgments}

We thank Gwynn J. Elfring and Saverio E. Spagnolie for the invitation to 
present preliminary results of this work at the Banff International Research 
Station ``Complex Fluids in Biological Systems'' workshop. We also thank Luca 
Martinetti for steady shear and oscillatory shear rheology measurements. 

\section*{References}

\bibliography{mybibfile}

\end{document}